\documentclass{emulateapj}
\usepackage{epsfig}
\usepackage{color}

\newcommand   {\about} {\mbox{$\sim$}}
\newcommand   {\mic}   {\mbox{$\mu$m}}

\newcommand   {\hh}    {\mbox{H$_2$}}

\newcommand   {\hcop}  {\mbox{HCO$^+$}}

\newcommand   {\nddd}  {\mbox{ND$_3$}}
\newcommand   {\nddh}  {\mbox{ND$_2$H}}

\newcommand   {\hho}   {\mbox{H$_2$O}}

\renewcommand {\deg}   {\mbox{$^\circ$}}

\newcommand   {\arcs}  {\mbox{$^{\prime\prime}$}}
\newcommand   {\pccm}  {\mbox{cm$^{-3}$}}
\newcommand   {\pscm}  {\mbox{cm$^{-2}$}}
\newcommand   {\kms}   {\mbox{km\,s$^{-1}$}}
\newcommand   {\ps}   {\mbox{s$^{-1}$}}
\renewcommand {\ga}    {\mbox{\rlap{\hbox{\lower5pt\hbox{$\sim$}}}\hbox{$>$}}}
\renewcommand {\la}    {\mbox{\rlap{\hbox{\lower5pt\hbox{$\sim$}}}\hbox{$<$}}}
\newcommand {\arcsper} {\mbox{\rlap{\hbox{\hbox{.}}}\hbox{$^{\prime\prime}$}}}

\newcommand  {\solar}  {\mbox{$_{\odot}$}}

\received{1 June 2011}
\accepted{23 June 2011}

\slugcomment{To appear in The Astrophysical Journal Letters}

\shortauthors{Lis et al.}

\shorttitle{Water Vapor at z=3.91}

\begin{document}

\title{Discovery of Water Vapor in the High-Redshift Quasar APM~08279+5255 at z=3.91}

\author{D.~C.~Lis$^1$, D.~A.~Neufeld$^2$, T.~G.~Phillips$^1$,
  M.~Gerin$^3$, and R.~Neri$^4$}


\altaffiltext{1}{California Institute of Technology, Cahill Center for
  Astronomy and Astrophysics 301-17, Pasadena, CA~91125;
  dcl@caltech.edu, tgp@submm.caltech.edu}

\altaffiltext{2}{Department of Physics and Astronomy, The Johns
  Hopkins University, 3400 North Charles Street, Baltimore, MD 21218;
  neufeld@jku.edu}

\altaffiltext{3}{LERMA, UMR 8112 du CNRS, Observatoire de Paris and
  Ecole Normale Superi\'{e}ure, 24 Rue Lhomond, 75231 Paris cedex 05,
  France; maryvonne.gerin@lra.ens.fr}

\altaffiltext{4}{IRAM -- Institut de Radio Astronomie Millimétrique,
  300 rue de la Piscine, 38406 Saint-Martin d’Hères, France;
  neri@iram.fr, cox@iram.fr}

\begin{abstract}

  We report a detection of the excited $2_{20}-2_{11}$ rotational
  transition of para-\hho\ in APM~08279+5255 using the IRAM Plateau de
  Bure interferometer. At $z=3.91$, this is the highest-redshift
  detection of interstellar water to date. From LVG modeling, we
  conclude that this transition is predominantly radiatively pumped
  and on its own does not provide a good estimate of the water
  abundance. However, additional water transitions are predicted to be
  detectable in this source, which would lead to an improved
  excitation model. We also present a sensitive upper limit for the HF
  $J=1-0$ absorption toward APM~08279+5255. While the face-on geometry
  of this source is not favorable for absorption studies, the lack of
  HF absorption is still puzzling and may be indicative of a lower
  fluorine abundance at $z=3.91$ compared with the Galactic ISM.
\end{abstract}


\section{Introduction}

Molecules such as CO or HCN have been commonly used as tracers of
molecular gas in high-redshift galaxies. However, recent observations
with the \emph{Herschel Space Observatory} \citep{pilbratt10} have
shown strong spectroscopic signatures from other light hydrides, such as
water, H$_2$O$^+$, or HF, in nearby active galaxies (e.g.,
\citealt{vanderwerf10}). These lines are blocked by the Earth's
atmosphere, but can be observed, redshifted, in distant galaxies using
the current millimeter and submillimeter facilities. For example,
\cite{omont11} have recently reported a detection of water in
J090302-014127B (SDP.17b) at $z = 2.30$.

One of the exciting recent results from HIFI \citep{degraauw10} is the
detection of widespread absorption in the fundamental $J=1-0$
rotational transition of hydrogen fluoride toward Galactic sources
\citep{neufeld10, phillips10, sonnen10, monje11}. Fluorine is the only
atom that reacts exothermically with \hh\ \citep{neufeld05,
  neufeld09}. The product of this reaction, HF, is thus easily formed
in regions where \hh\ is present and its very strong chemical bond
makes this molecule relatively insensitive to UV photodissociation. As
a result, HF is the main reservoir of fluorine in the interstellar
medium (ISM), with a fractional abundance of $\about 3.5 \times
10^{-8}$ relative to \hh\ typically measured in diffuse molecular
clouds within the Galaxy \citep{neufeld10, sonnen10, monje11}.

Interstellar HF was first detected by \cite{neufeld97} with the
Infrared Space Observatory (ISO). The $J=2-1$ rotational transition
was observed in absorption toward Sagittarius~B2, at a low spectral
resolution using the Long-Wavelength Spectrometer (LWS). The HIFI
instrument allows for the first time observations of the fundamental
rotational transition of HF at 1.232476~THz to be carried out, at high
spectral resolution. Given the very large Einstein A coefficient
($2.423 \times 10^{-2}$~\ps; critical density $\about 3 \times
10^9$~\pccm), this transition is generally observed in absorption
against dust continuum background. Only extremely dense regions with
strong IR radiation field could possibly generate enough collisional
or radiative excitation to yield an HF feature with a positive
frequency-integrated flux.\footnote{HF \emph{emission} has recently
  been reported in the extreme environment of the nearby ultraluminous
  galaxy Markarian~231 in Herschel/SPIRE observations at low spectral
  resolution \citep{vanderwerf10}.} The HIFI observations corroborate
the theoretical prediction that HF will be the dominant reservoir of
interstellar fluorine under a wide range of interstellar conditions.
The HF $J=1-0$ transition promises to be a excellent probe of the
kinematics of, and depletion within, absorbing material along the line
of sight toward bright continuum sources, and one that is
uncomplicated by the collisionally-excited line emission that is
usually present in the spectra of other gas tracers. As suggested by
\cite{neufeld05}, redshifted HF $J=1-0$ absorption may thus prove to
be an excellent tracer of the interstellar medium in the high-redshift
Universe, although only the gas reservoir in front of a bright
continuum background can be studied by means of the HF absorption
spectroscopy.

Water is another interstellar molecule of key importance in
astrophysical environments, being strongly depleted on dust grains in
cold gas, but abundant in warm regions influenced by energetic process
associated with star formation (see \citealt{vandishoeck11} and
references therein). The excited $2_{20}-2_{11}$ transition of p-\hho,
with a lower level energy of 137~K, has a frequency of 1.228788~THz
and can be observed simultaneously with the $J=1-0$ transition of HF
in high-redshift systems. Consequently, we have searched for the HF
$J=1-0$ and \hho\ $2_{20}-2_{11}$ transitions, redshifted down to
251~GHz, in APM~082791+5255 using the IRAM Plateau de Bure
interferometer.

The broad absorption line (BAL) quasar APM~082791+5255 at
\emph{z}=3.9118, with a true bolometric luminosity of $(0.7-3) \times
10^{14}$ L$_\odot$ , is one of the most luminous objects in the
Universe \citep{downes99}. CO lines up to $J=11-10$ have been detected
using the IRAM~30-m telescope. IRAM PdBI high spatial resolution
observations of the CO $J=4-3$ and $9-8$ lines, and of the 1.4~mm dust
continuum have been presented by \cite{weis07}. The line fluxes in the
CO ladder and the dust continuum fluxes are well fit by a
two-component model that invokes a “cold” component at 65~K with a
high density of $n$(H$_2$) = $1 \times 10^5$~cm$^{-3}$, and a “warm”,
$\sim 220$~K, component with a density of $1 \times 10^4$~cm$^{−3}$.
Wei\ss\ et al. argue that the molecular lines and the dust continuum
emission arise from a very compact ($r \simeq 100-300$~pc), highly
gravitationally magnified ($m \simeq 60-110$) region surrounding the
central AGN. Part of the difference relative to other high-\emph{z} QSOs may
therefore be due to the configuration of the gravitational lens, which
gives us a high-magnification zoom right into the central 200-pc
radius of APM~08279+5255 where IR pumping plays a significant role for
the excitation of the molecular lines.

High-angular resolution ($0.3\arcs$) VLA observations of the CO
$J=1-0$ emission in APM~08297+5255 \citep{riechers09} reveal that the
molecular emission originates in two compact peaks separated by \la
0\arcsper 4 and is virtually co-spatial with the optical/near infrared
continuum emission of the central active galactic nucleus (AGN). This
morphological similarity again indicates that the molecular gas is located
in a compact region, close to the AGN. \cite{riechers09} present a
revised gravitational lens model of APM~08297+5255, which indicates
a magnification by only a factor of 4, in contrast to much higher
magnification factors of \about 100 suggested in earlier studies. Their
model suggests that the CO emission originates from a \about 550~pc
radius circumnuclear disk viewed at an inclination angle of \la
25\deg, or nearly face-on. The total molecular mass is then $1.3 \times
10^{11}$~M\solar. 

\cite{weis07} first pointed out the importance of infrared pumping
for the excitation of HCN in APM~08279+5255. Subsequent observations of
\cite{riechers10} reveal surprisingly strong $J=6-5$ emission of HCN,
HNC, and \hcop\ in the host galaxy, providing additional evidence that
these transitions are not collisionally excited. \cite{riechers10}
argue that the high rotational lines of HCN can be explained by
infrared pumping at moderate opacities in a \about 220~K warm gas and
dust component. These findings are consistent with the overall picture
in which the bulk of the gas and dust is situated in a compact,
nuclear starburst, where both the AGN and star formation contribute to
the heating.

Prior to the observations reported here, water had not been detected
in APM~08279+5255. However, \cite{wagg06} give an upper limit of 
0.7~Jy\,km\,s$^{-1}$ ($3 \sigma$) for the ground state $1_{10}-1_{01}$
ortho-\hho\ line.

\section{Observations}\label{observations}

Observations of APM~ 08279+5255 presented here were carried out on
2010 June 22, September 21--22, and December 15, using the Plateau de
Bure interferometer. Visibilities were obtained in the CD set of
configurations of the six-element array in June and December and with
a four-element subarray in September, totalling 4.9 hr of on-source
observations.

Data reduction and calibration were carried out using the GILDAS
software package in the standard antenna based mode. The passband
calibration was measured on 3C454.3, and amplitude and phase
calibration were made on 0749+540, 0836+716 and 0917+449. The absolute
flux calibration, performed using MWC349 as the primary
calibrator (2.55~Jy at 250~GHz), is accurate to within 10\%. Point
source sensitivities of 4.5\,mJy\,beam$^{-1}$ were obtained in channels of
20\,MHz, consistent with the measured system temperatures
(200--300~K). The conversion factor from flux density to brightness
temperature in the $1.6\arcs \times 1.3\arcs$ (PA=22\deg) synthesized
beam is 9.1~K\,(Jy\,beam$^{-1}$)$^{-1}$.

\begin{figure}[tb]
\centering
\includegraphics[width=0.95\columnwidth,angle=0]{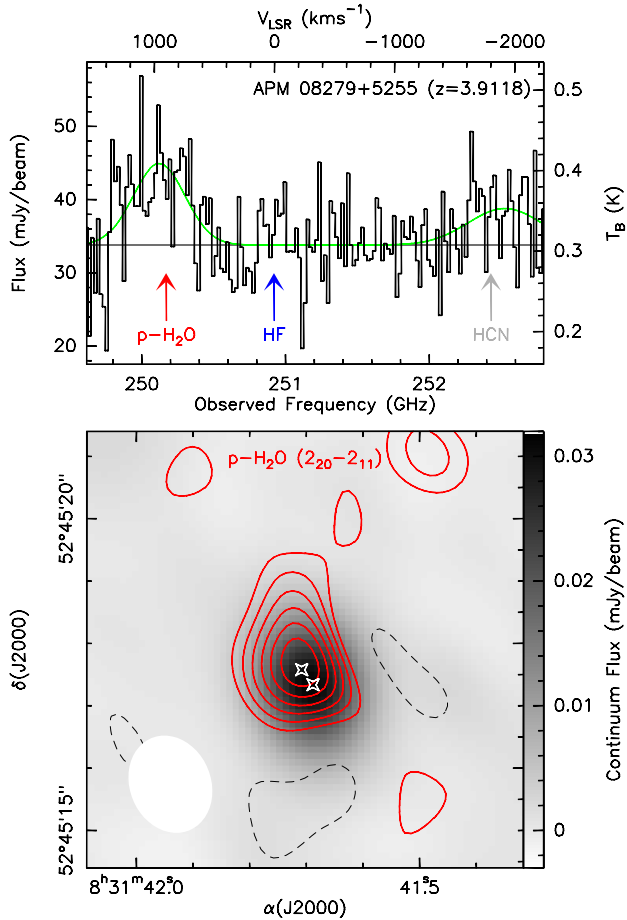}
\caption{(Top) Integrated spectrum of APM~08279+5255. Vertical arrows
  mark redshifted frequencies of para-\hho\ $2_{20}-2_{11}$, HF
  $J=1-0$, and HCN $J=14-13$, assuming z=3.9118. Velocity scale is
  with respect to the HF $J=1-0$ frequency. HF and HCN lines are not
  detected. (Bottom) Distribution of the velocity-integrated
  para-\hho\ $2_{20}-2_{11}$ line intensity in APM~08279+5255 (red
  contours) superposed on a grayscale image of the 1.23~THz (rest
  frame) dust continuum emission. Contour levels are -2, 2, 3, 4, 5,
  6, 7 times the rms of 0.8~Jy\,\kms\,beam$^{-1}$. White symbols mark
  the locations of sources A and B \citep{riechers09}. Synthesized
  beam is show as a white ellipse in the lower-left corner.}
\label{fig:image}
\end{figure}

\section{Results}\label{results}

Figure~\ref{fig:image} (upper panel) shows a spectrum of
APM~08279+5255 near the rest frame frequency of the HF $J=1-0$
transition, integrated over the PdBI image, which also covers
frequencies of the para-\hho\ $2_{20}-2_{11}$ and HCN $J=14-13$
transitions, in addition to HF. The continuum is detected with a high
SNR. The integrated flux density, computed from emission free
channels, is $34 \pm 0.55$~mJy, consistent with the previous
measurements of the source SED \citep{weis07}. No HF absorption is
seen, with a 3$\sigma$ upper limit of 1.5~Jy\,\kms, assuming a FWHM
line width of 500~\kms, as implied by earlier CO observations.

The integrated line and continuum fluxes given above impose a
3$\sigma$ upper limit of 0.092 for the velocity-averaged HF $J = 1 -
0$ optical depth (velocity-integrated optical depth $\tau dv =
46$~\kms). The corresponding column density of cold HF in front of the
continuum source can then be computed using eq. (3) of
\cite{neufeld10} to be $N({\rm HF}) < 1.1 \times 10^{14}$~\pscm. Given
the typical Galactic HF/\hh\ abundance ratio of $\about 3.5 \times
10^{-8}$, this value would imply an average \hh\ column density
$\about 3 \times 10^{21}$~\pscm\ lying in front of the continuum
source in APM~08279+5255. This value is three orders of magnitude
below the beam averaged \hh\ column density inferred from the dust
continuum flux observed toward the source.\footnote{In their
  two-component model for the dust continuum radiation,
  \citealt{weis07} derived estimates of $\about 2.8 \times 10^9
  m^{-1}$ M\solar\ and 680 pc, respectively, for the dust mass and
  magnified radius for APM 08279+5255, where $m$ is the lens
  magnification. These values imply an average \hh\ column density of
  $1.2 \times 10^{25}$~\pscm\ for an assumed dust-to-gas mass ratio of
  100. Approximately 50\% of this gas should be in front of the
  continuum source.}

The para-\hho\ $2_{20}-2_{11}$ line is clearly detected with the
integrated line flux density of $6.7\pm 0.8$~Jy\,\kms. A Gaussian fit
gives a line width of $510 \pm 70$~\kms, consistent with that of CO.
Figure~\ref{fig:image} (lower panel) shows spatial distribution of the
para-\hho\ $2_{20}-2_{11}$ emission (red contours) superposed on a
grayscale image of the dust continuum. The line and continuum emission
peak toward sources A and B of \cite{riechers09}. The small offset
between the \hho\ and continuum emission is not significant at the
spatial resolution of the present observations. Implication for water
excitation in APM~08279+5255 are discussed below.

Some excess emission above the continuum level is seen near the
frequency of the HCN~$J=14-13$ line, however, the result does not
constitute a detection at the sensitivity limit of the present observations.

\section{Water Excitation}\label{discussion}

In modeling the water line flux observed from APM~08279+5255, we have
computed the CO and H$_2$O line luminosities expected for an
isothermal, constant density medium. We solved the equations of
statistical equilibrium for the H$_2$O and CO level populations,
making the large velocity gradient (LVG) approximation and treating
the effects of radiative trapping with an escape probability method.
We adopted the rate coefficients of \cite{yang10} and \cite{faure08},
respectively, for the excitation of CO and H$_2$O in collisions with
H$_2$, and we assumed an ortho-to-para ratio (OPR) of 3 for both H$_2$
and H$_2$O. Following \cite{weis07}, we neglect any effects of dust
extinction upon the emergent CO and \hho\ line fluxes; although the
dust optical depths at THz frequencies raise the possibility that such
effects could be significant, their importance depends strongly on the
geometry of the source and the spatial relationship between the warm
dust and the molecular emission region. In Galactic hot cores with
\hh\ column densities comparable to that in APM~08279+5255, the
para-\hho\ $2_{20}-2_{11}$ line can be seen with net-emission flux
(Orion KL, NGC6334I), or in absorption (Sagittarius~B2), depending on
the specific source geometry. A mixture of such regions may contribute
to the observed spectrum of APM~08279+5525, leading to partial
cancellation of the emergent line flux. Using the relative strengths
of the multiple CO transitions observed by \cite{weis07} to constrain
the gas temperature, density, and velocity gradient, we thereby
obtained as best fit parameters the values $T=10^{2.35}$~K, $n({\rm
  H}_2) = 10^{4.2}\, \rm cm^{-3}$, and $dv/dz= 1.4 \times 10^{5}
n({\rm CO}) / n({\rm H}_2) \rm \, km \, s^{-1} \, pc^{-1}$,
respectively. These parameters are very close to those obtained
previously by \cite{weis07} in their single component model for the CO
emission detected from this source.

Adopting the same parameters for the water emitting region, we have
computed the para-H$_2$O~$2_{20} - 2_{11}$/CO~$J = 11-10$ line flux
ratio as a function of the assumed $n({\rm H_2O})/n({\rm CO})$
abundance ratio. In the case of H$_2$O, the pumping of rotational
transitions by far-infrared continuum radiation can strongly affect
the predicted line fluxes. The importance of radiative pumping in this
source has been discussed previously by \cite{riechers10} for the case
of HCN, although, in that case, pumping takes place through a
low-lying {\it vibrational} band. Pumping through pure {\it
  rotational} transitions is relatively much more important for an
asymmetric top molecule like water, because such molecules possess a
more complex energy level structure than the simple ladder shown by
spinless linear or diatomic molecules (such as HCN and CO);
furthermore, the lowest vibrational band of water lies at a
considerably shorter wavelength ($\sim 6.3 \, \mu\rm m$) than that of
HCN ($\sim 14.7 \, \mu\rm m$), where the continuum radiation is
considerably weaker. The dominance of radiative pumping in rotational
transitions of water vapor was also discussed by \cite{gonzalez10}, in
their recent analysis of the water line emission observed by {\it
  Herschel} toward the starburst galaxy Mrk~231. Under conditions
where radiative pumping is dominant, the water line fluxes are almost
independent of the gas temperature and density.

\begin{figure*}[tb]
\centering
\includegraphics[width=0.70\textwidth,angle=90]{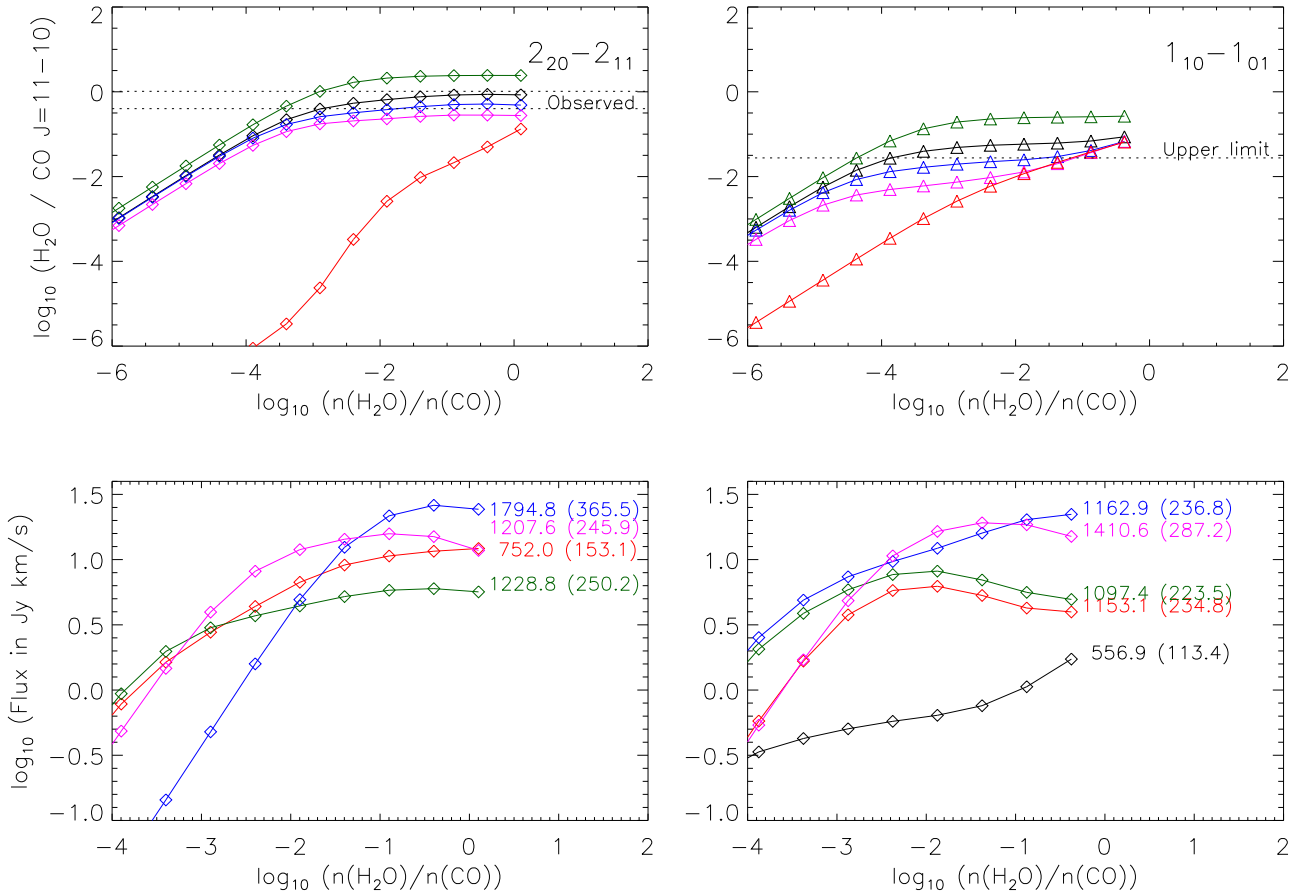}
\caption{Molecular line ratios predicted by LVG models of
  APM~08279+5255, as described in the text. Labels in the lower two
  panels give the frequencies in GHz (wavelengths in \mic) of water
  transitions that are predicted to be strong and potentially
  detectable.}
\label{fig:model}
\end{figure*}

Following \cite{riechers10}, we define $\rm IR_{ff}$ as the sky
covering factor of the infrared continuum source at the location of
the molecular emission region; the mean intensity is then given by
modified blackbody of the form ${\rm IR_{ff}}(1 - e^{-\tau}) B_\nu
({\rm 220\,K}) $, where $B_\nu(T)$ is the Planck function and
$\tau=1.05\, (\nu/\rm 1\,THz)^{2}$. In Figure~\ref{fig:model} (upper left panel),
we present the predicted para-H$_2$O $2_{20} - 2_{11}$ / CO $J =
11-10$ line flux ratio, as a function of $n({\rm H_2O})/n({\rm CO})$
and for several different values of $\rm IR_{ff}$: 0 (red), 0.1
(magenta), 0.25 (blue), 0.5 (black), and 1.0 (green). The results
shown in Figure~\ref{fig:model} clearly indicate the dominant role of radiative
pumping; a detailed analysis indicates the importance of the $1_{11}
\rightarrow 2_{20}$ (rest frame 2.97~THz) transition in directly
pumping the $2_{20}$ state of para-water, along with the $2_{02} -
3_{31}$ (rest frame 6.45~THz) transition, which pumps the $3_{31}$
state; the latter can decay subsequently to $2_{20}$. Dotted
horizontal lines indicate the para-H$_2$O $2_{20} - 2_{11}$ / CO $J =
11-10$ ratio measured in APM08279+5255 and its uncertainty.

The upper right panel of Figure~\ref{fig:model} shows entirely
analogous results for the $1_{10}-1_{01}$ transition of ortho-water,
with the horizontal dotted line indicating the 3 $\sigma$ upper limit
obtained by \cite{wagg06}. A comparison of the results shown in the
upper panels of Figure~\ref{fig:model} indicates that a limited range
of parameters is permitted by the measured value of the $2_{20} -
2_{11}$ line flux and the upper limit on $1_{10} - 1_{01}$. For the
H$_2$O OPR of 3 assumed here, acceptable fits are obtained for $\rm
IR_{ff} \sim 0.25$ and $n({\rm H_2O})/n({\rm CO})$ in the range $\sim
0.003 - 0.03$, although the range of acceptable parameters would
obviously broaden if OPR values smaller than 3 were permitted. In the
regime of interest, the expected $2_{20}-2_{11}$ line fluxes depend
only weakly upon the H$_2$O abundance, the most important pumping
transitions being optically-thick; thus, the exact range of acceptable
values for $n({\rm H_2O})/n({\rm CO})$ depends strongly upon our
estimate of the likely error in the measured line flux. Nevertheless,
the $n({\rm H_2O})/n({\rm CO})$ ratio inferred for an assumed OPR of 3
is apparently smaller than that typically measured ($\sim 0.1 - 1$) in
hot core regions within our Galaxy (e.g.~\cite{boonman03}).
Observations of additional transitions will be needed to constrain the
water OPR and abundance better. The lower panels of
Figure~\ref{fig:model} present results for several other transitions
that are potentially detectable from ground-based observatories, some
of which are more strongly dependent upon the water abundance. The
results shown in these panels were all obtained for $\rm IR_{ff} =
0.25$ and OPR=3, and the labels indicate the rest frequencies and --
in parentheses -- redshifted frequencies in GHz. Our results for
transitions of para- and ortho-water appear, respectively, in the left
and right panels.

We note that our LVG solution for the \hho\ emission in APM~08279+5525
not only matches that of \cite{weis07}, using independent data, but
the deduced velocity gradient $dv/dz$ for the \hho\ emitting region is
close to the expected virial value, as defined for example by eq. (5)
of \cite{greve09}, $K_{vir} \about 4$ (i.e. the dense gas emitting in
the \hho\ line emission is near virial equillibrium).

\section{Discussion}

The absence of detectable HF~$J=1-0$ absorption in APM~08279+5255 is
unexpected, given the low column density of HF required to produce
measurable absorption. An important caveat in this analysis is the
assumption of a Galactic HF/\hh\ ratio that, in turn, is related to
the elemental abundance of fluorine. Fluorine nucleosynthesis---and
thus the evolution of the fluorine abundance in cosmic time---remains
poorly understood, with production in AGB stars (e.g.
\citealt{cristallo09}), in Wolf-Rayet stars \citep{meynet00} and
neutrino-induced nucleosynthesis in Type II supernovae
\citep{woosley98} all proposed as possible mechanisms. While the
face-on geometry of APM~08279+5255 is not favorable for absorption
studies, the lack of HF absorption is still puzzling and may be
indicative of a lower fluorine abundance in this source compared with
the Galactic ISM. Nevertheless, HF absorption may still prove to be a
good tracer of \hh\ in high-redshift sources and additional
observations of objects with different geometries, over a wide range
of redshifts, are urgently needed.

Our LVG models indicate that the para-\hho\ $2_{20} - 2_{11}$
transition in APM~8279+5255 is predominantly radiatively pumped.
\cite{omont11} has reached similar conclusions regarding water
excitation in J090302-014127B (SDP.17b) at $z = 2.30$. The para-\hho\
$2_{20} - 2_{11}$ line intensity in APM~8279+5525 is sensitive to the
details of the excitation model. Consequently, observations of this
single transition do not provide a good estimate of the water
abundance. However, our LVG models suggests that many additional water
lines should be detectable with the current millimeter-wave
facilities. The transitions that are expected to be the strongest (see
Fig.~2) are: $3_{21}-3_{12}$ (rest frame frequency 1162.2~GHz),
$4_{22}-4_{13}$ (1207.6~GHz), $2_{11}-2_{02}$ (752.0~GHz), as well as
two very high-energy transitions: $6_{24}-6_{15}$ (1794.8~GHz), and
$5_{23}-5_{14}$ (1410.6~GHz), which are sensitive to the gas density.
With multi-line observations, the excitation conditions and the water
abundance will be much better constrained. This excitation scenario
can further be tested with observations of the 2.97~THz pumping
transition, which is expected to appear in absorption.

\acknowledgments 

Based on observations carried out with the IRAM Plateau de Bure
interferometer. IRAM is supported by INSU/CNRS (France), MPG (Germany)
and IGN (Spain). This research has been supported by the National
Science Foundation grant AST-0540882 to the Caltech Submillimeter
Observatory. We thank Pierre Cox for allocating Director's
Discretionary Time to allow these observations to be carried out and
an anonymous referee for constructive and helpful comments.

\end{document}